\documentclass[aps,prb,twocolumn,groupedaddress,showpacs]{revtex4}

\usepackage{amssymb,amsmath}

\newcommand{\hc}{\mathrm{ h. c. }} 
\newcommand{\subcomma}{\! , \,}
\newcommand{\calr}{\mathbf{r}}
\newcommand{\calJ}{\mathcal{J}}
\newcommand{\dos}{\mathcal{N}}
\newcommand{\mytrace}{\mathrm{Tr}}
\newcommand{\fraku}{\mathfrak{u}}

\newcommand{\br}{\mathbf{r}}
\newcommand{\bh}[1]{\mathbf{\hat{#1}}}

\begin{document}
\title{Exact summation of vertex corrections to the penetration depth\\ in $d$-wave superconductors}
\author{A. Iyengar}
\author{M. Franz}
\affiliation{Department of Physics and Astronomy, University of British Columbia, Vancouver, British Columbia, Canada V6T 1Z1}

\date{\today}

\begin{abstract}
A variety of experiments suggest that in the cuprates, the low-energy superconducting quasiparticles undergo forward
scattering from extended impurity potentials.  We argue that when such potentials dominate the scattering, the penetration
depth may be computed in a simple zero-angle scattering approximation (ZSA), in which 
the vertex corrections to the Meissner effect may be summed exactly.  We find a remarkably simple relationship 
between the normal fluid density and the quasiparticle density of states of the disordered system which holds 
for every realization of the disorder.
We expect this result to be relevant to the $ab$-plane penetration depth in high-purity single crystals of underdoped 
YBCO. 
\end{abstract}


\maketitle


Measurements of the in-plane penetration depth $\lambda$ have been central to elucidating the nature of 
superconductivity in the high-temperature copper-oxide superconductors. Equally important has been the theoretical 
understanding of impurity scattering in $d$-wave superconductors required to interpret these measurements.
The temperature dependence of $\lambda(T)$
provided the first strong evidence\cite{HardyBonnDwave}  for unconventional pairing symmetry. 
Subsequently, the effect on $\lambda(T)$ of isotropic elastic 
scattering from point-like in-plane impurities in a superconductor with $d_{x^2 - y^2}$ pairing symmetry was  
investigated.\cite{HirschfeldPutikkaScalapino} This picture of impurity scattering 
accounted for the observed crossover from a quadratic 
low-temperature behavior in $\lambda^{-2}(T)$ to the $T$-linear behavior at higher temperatures characteristic of 
a gap function with nodes.

More recently, it has been suggested in different experimental contexts that quasiparticles near the Fermi
surface undergo \emph{forward} scattering. 
Forward scattering from impurities has been invoked \cite{AbrahamsVarma} 
to account for a component of the single-particle scattering rate observed in angle-resolved 
photoemission (ARPES) spectra which apparently contributes negligibly to the normal state resistivity.
Such scattering may arise from a smoothly varying in-plane potential due to the poorly screened\cite{PanTheory}
Coulomb fields of disordered dopant oxygen ions residing between the CuO$_2$ layers. This potential may account for 
the nanoscale electronic inhomogeneity observed in scanning-tunneling spectroscopy (STS) studies of 
optimally to overdoped Bi$_2$Sr$_2$CaCu$_2$O$_{8+x}$.\cite{PanNature} A flurry of recent 
activity\cite{HirschfeldFTSTS,MarkiewiczCondmat,HirschfeldJun04,HirschfeldSept04,HirschfeldOct04} is aimed in part at 
synthesizing the evidence for forward scattering in this compound 
obtained from ARPES, STS, and thermal and electrical conductivities. 

At the same time, measurements of the microwave conductivity $\sigma(\omega,T)$ in YBCO present their own puzzles.
High quality YBa$_2$Cu$_3$O$_{6.993}$ crystals have shown a Drude-like
phenomenology in the conductivity at low frequencies and temperatures.\cite{Hosseini} In a nodal quasiparticle 
picture of low-energy charge transport\cite{Berlinsky}, this implies
significant frequency-independent part of the scattering rate, inconsistent with scattering by point defects.
It has been shown\cite{DurstLeeLinearDefects} that extended linear defects such as twin boundary remnants 
can produce such behavior. However, precise bolometric measurements\cite{TurnerPRL} subsequently have shown deviations
from Drude behavior at the lowest temperatures and frequencies as well as an unexpected $\omega/(T+T_0)$ scaling. 
Violation of the Wiedemann-Franz law and the universal conductivity limit\cite{LeeLimit} in cuprates
have also fueled theoretical interest in quasiparticle self-energy\cite{SheehyDOS} and 
transport\cite{DurstLeeVC,SheehyExtendedRSRG} in the presence of extended impurities.

Finally, extended disorder potentials have surfaced in a recent attempt\cite{UnifiedABC} to explain in a unified theory
the doping and temperature dependence of both the $c$-axis and $ab$-plane penetration depth in underdoped YBCO.
The theory models the incoherent interlayer hopping by scattering in-plane states by momenta of order $\Lambda$ 
as they tunnel between the layers.  Good agreement with recent $c$-axis data has been obtained\cite{BrounPRL}
with $\hbar / \Lambda = 120$\AA, or about $25$ lattice spacings.
The effect of such interlayer disorder on in-plane transport is 
similar to that of in-plane scattering potentials which extend over distances of order $\hbar/\Lambda$. Presumably, such
potentials arise from the aforementioned disorder of interlayer oxygen dopants, which, unlike BSCCO, may dominate the 
scattering rate due to the very low cation disorder in high-purity YBCO crystals.

In this paper, we argue that the nodal structure of a $d$-wave superconductor presents an unusual situation in which 
even mildly extended potentials lead to strong forward scattering.  The simplest approximation allowing 
for a sensible calculation of the superfluid density is  
the zero-angle scattering approximation (ZSA):
the impurity potential is unable to
modify upon scattering the charge current carried by a quasiparticle.  Extended potentials induce 
elastic scattering by small momenta, implying that, except 
via rare internodal scattering events, nodal quasiparticles scatter exclusively within the same node.  
However, the quasiparticle
velocity relevant to charge current varies much more slowly in the nodal region than does the energy, as it arises 
from the \emph{bare} bandstructure.  The ZSA is thus expected to be valid when 
extended potentials dominate the scattering rate.  This is likely the situation in high-purity YBCO single crystals, 
in which smooth potentials due to interlayer dopants dominate the scattering.  

The main result
presented here is that under such circumstances the normal fluid density $n_n$ 
is connected to the single-particle density of 
states per site per spin $\dos(E)$ by the simple relation
\begin{equation}
\frac{n_n}{m^*} = \left(\frac{v_{F}}{a}\right)^2 \int dE\: \dos(E)[-f'(E)],
\label{mainresult}
\end{equation}
where $v_F$ is the magnitude of the quasiparticle velocity at the node and $f$ is the Fermi-Dirac distribution.
Since the electron density is the sum of the normal- and superfluid densities, this quantity dictates the 
temperature dependence of the penetration depth. 
It is remarkable that \eqref{mainresult}, which is a trivial result for the clean superconductor, continues to 
hold in the presence of extended disorder potentials, albeit with $\dos(E)$ renormalized by the disorder.
Since it is valid for every realization of the disorder, 
it circumvents the considerable complications of vertex corrections,\cite{DurstLeeVC}
which arise upon disorder-averaging two-particle correlation functions.
Furthermore, \eqref{mainresult} does not require any particular model of the disorder 
(e.g. Born limit, ladder approximation,) though it reveals nothing about 
$\dos(E)$ itself.
Thus if the ZSA is valid, 
\eqref{mainresult} is a powerful consequence
allowing a model-independent interpretation of the normal fluid density.

The limitation of this approximation
is that, as in the clean superconductor,  the dissipative part of conductivity 
\endnote{Hereafter we use $\sigma_1(\omega)$ and ``conductivity'' to denote only the normal fluid component.}
$\sigma_1(\omega)$ becomes a 
$\delta$-function at zero frequency. This is to be expected since zero-angle scattering cannot degrade the 
charge current. However, as in a clean superconductor at $T\ne 0$,  
a fraction of the superfluid is converted into normal fluid,
which may be reliably computed despite the fact that the distribution of $\sigma_1(\omega)$ is unphysical.
In the ZSA,
this normal fluid behaves as a perfect metal, having infinite d.c. conductivity but making no contribution to the Meissner 
effect.  In reality, small-angle (and rare internodal)
scattering degrades this perfect metal and results in a dissipative conductivity 
peak of finite width.  However, the ZSA can provide no information about the frequency distribution of
$\sigma_1(\omega)$. 
We assume that such scattering is sufficiently weak that it
primarily affects only the frequency distribution of $\sigma_1(\omega)$ and not the integral, 
so that Eq.~\eqref{mainresult} remains valid.

We consider the general Hamiltonian $H = H_K + H_{sc}$, with
\begin{widetext}
\begin{equation}
\begin{split}
H_K = \mathop{\sum_{\br\subcomma\sigma}}&\left[ 
\frac{1}{2}(V_{\br} - \mu) \: c_{\br\subcomma\sigma}^{\dagger} c_{\br\subcomma\sigma}^{}
-t\,  (c_{\br + \bh{x}a\subcomma\sigma}^{\dagger} c_{\br\subcomma\sigma}^{} + 
c_{\br + \bh{y}a\subcomma\sigma}^{\dagger} c_{\br\subcomma\sigma}^{})
-t^{\perp}_{\br}\:  c_{\br + \bh{z}d\subcomma\sigma}^{\dagger} c_{\br\subcomma\sigma}^{}
\right] \quad + \hc \: ,
\\
H_{sc} = \frac{1}{2}\mathop{\sum_{\br}}&\left[
\Delta_{\br + \bh{x}a/2} 
(c^{\dagger}_{\br\subcomma\uparrow} c^{\dagger}_{\br+\bh{x}a\subcomma\downarrow} 
+ c^{\dagger}_{\br + \bh{x}a\subcomma\uparrow} c^{\dagger}_{\br\subcomma\downarrow})
- 
\Delta_{\br + \bh{y}a/2} 
(c^{\dagger}_{\br\subcomma\uparrow} c^{\dagger}_{\br+\bh{y}a\subcomma\downarrow} 
+ c^{\dagger}_{\br + \bh{y}a\subcomma\uparrow} c^{\dagger}_{\br\subcomma\downarrow})
\right] \quad + \hc 
\end{split}
\label{Hamiltonian}
\end{equation}
\end{widetext}
It describes, 
at the mean-field level, a $d$-wave superconductor on square-lattice layers with in-plane lattice constant $a$ and 
interlayer spacing $d$.
The interlayer hopping $t^{\perp}_{\br}$ (which may model the incoherent $c$-axis transport as in 
Ref. \onlinecite{UnifiedABC}) and the on-site potential $V_{\br}$ are assumed to be weak relative to 
$t$ and have a zero average value. The order parameter $\Delta_{\br}$ is defined on lattice bonds and 
may also vary weakly (compared to its average value) due to the disorder potentials.
These quantities 
are assumed to vary slowly along the in-plane directions
over the length scale $\hbar / \Lambda$ but may vary randomly from layer to layer. 
 
It will be convenient to work with the Nambu spinor
$\Psi_{\br} \equiv (
c_{\br\subcomma\uparrow}^{},c_{\br\subcomma\downarrow}^{\dagger})^T$. 
The Hamiltonian \eqref{Hamiltonian} may be diagonalized by a 
Bogliubov transformation to 
$H = E_0 + \mathop{\sum_{n\subcomma\sigma}} \epsilon_n \,
\gamma_{n\subcomma\sigma}^{\dagger}\gamma_{n\subcomma\sigma}^{} $
where $E_0$ is the ground state energy and $\epsilon_n$ labels the (positive) quasiparticle excitation energies. 
The $\gamma$ operators obey the usual fermion commutation relations up to a renormalization constant: 
$\{ \gamma_{\br\subcomma\sigma}^{},\gamma_{\br'\subcomma\sigma'}^{\dagger} \} 
= N \delta_{\br\subcomma \br'} \delta_{\sigma\subcomma\sigma'}$
where $N$ in a finite system is the number of lattice sites.
The transformation can be compactly represented using a two-component 
wavefunction 
$\fraku^{(n)} (\br) = (u_n(\br),v_n(\br))^T$.
Introducing the antisymmetric tensor $\varepsilon$, we write
(with the sums over tensor indices implicit hereafter,)
$\gamma_{n\subcomma\uparrow}^{\dagger} = \mathop{\sum_{\br}} 
(\Psi_{\br}^{\dagger})_{\alpha} \fraku^{(n)}_{\alpha}(\br)$
and 
$\gamma_{n\subcomma\downarrow}^{} = \mathop{\sum_{\br}} 
\varepsilon_{\alpha\alpha'}
(\Psi_{\br}^{\dagger} )_{\alpha'}
\fraku^{(n)\,*}_{\alpha}(\br)$.
The wavefunctions satisfy the orthogonality relations 
\begin{equation}
\begin{split}
\frac{1}{N}\mathop{\sum_{\br}}& 
\fraku^{(n)}_{\alpha}(\br)
\fraku^{(n')\:*}_{\alpha}(\br) = \delta_{n\subcomma n'} ,
\\
\frac{1}{N}\mathop{\sum_{\br}}& \varepsilon_{\alpha\beta}
\fraku^{(n)}_{\alpha}(\br)
\fraku^{(n')}_{\beta}(\br) = 0 .
\end{split}\label{frakunitarity}
\end{equation}



With this notation, the in-plane components of current are
$\calJ_{x(y)}(\calr) \equiv i t 
(\Psi_{\calr+\bh{x}(\bh{y})a}^{\dagger} \Psi_{\calr}^{} -
\Psi_{\calr}^{\dagger} \Psi_{\calr+\bh{x}(\bh{y})a}^{}
)
$.
In the disordered system, $n_n/m^* = \Pi_{xx}(0)$ where the long-wavelength polarization 
function is 
$
\Pi_{\mu\nu} (i\omega) \equiv \frac{1}{N}\mathop{\sum_{\calr\subcomma\calr'}}
\int_0^{\beta} d\tau\: e^{i\omega \tau}
\left< \mathcal{T}_{\tau} \calJ_{\mu}(\calr,\tau) \, \calJ_{\nu}(\calr',0) \right> 
$.
It is also useful to introduce a vertex function
which describes the coupling of long-wavelength radiation to electrons at lattice sites $\br_1$ and $\br_2$,
\begin{equation}
\lambda_{x(y)}(\br_1, \br_2) \equiv  
i t (\delta_{\br_1\subcomma \br_2 + \bh{x}(\bh{y})a}
- \delta_{\br_2\subcomma \br_1 + \bh{x}(\bh{y})a}) .
\label{lambda}
\end{equation}
Importantly, the Hamiltonian \eqref{Hamiltonian} describes a non-interacting system, 
and we may apply Wick's theorem to evaluate $\Pi$. 
We assume that $\left< \calJ \right> = 0$  in the unperturbed system, in which case
$\Pi$ may be expressed 
\begin{align}
\begin{split}
\Pi_{xx}(i\omega) &= 
- \frac{1}{N} \mathop{\sum_{\br_1 \subcomma \br_2}} \mathop{ \sum_{\br_1' \subcomma \br_2'}}
\lambda_x(\br_1,\br_2)  
\lambda_x(\br_1',\br_2') 
\\
T\mathop{\sum_{i \nu}}
&\mytrace \left[ G(\br_2, \br_1'; i\nu + i\omega) G(\br_2',\br_1; i\nu)  \right]
\end{split}
\label{Pi}
\end{align}
in terms of the matrix Green function  
$ 
G(\br,\br'; i\nu) \equiv - \int_0^{\beta} d\tau\: e^{i\nu\tau} 
<\!\!\! 
\mathcal{T}_{\tau} 
\Psi_{\br}^{}(\tau)\Psi_{\br'}^{\dagger}(0)
\!\!\!>
$. We will make use of the spectral function
$
A(\br,\br';\omega) \equiv 
(2\pi i)^{-1}
[G^A(\br,\br';\omega) - G^R(\br,\br';\omega)] 
$ 
which has the representation
\begin{align}
A(\br,\br';\omega)_{\alpha \beta} &= \frac{1}{N} \mathop{\sum_{n}}
\left[
\delta(\omega - \epsilon_n) 
\fraku^{(n)}_{\beta}(\br) \fraku^{(n)\:*}_{\alpha}(\br')
\right. \\ \nonumber &\left. + \quad
\delta(\omega + \epsilon_n) 
\varepsilon_{\alpha\alpha'} \varepsilon_{\beta\beta'}
\fraku^{(n)}_{\alpha'}(\br') \fraku^{(n)\:*}_{\beta'}(\br)
\right] .
\end{align} 


We now state the ZSA precisely, establish its validity, and derive \eqref{mainresult}.
We define
\begin{align}
Q_{x}(\br_1,\br_2;\omega) &\equiv
\mathop{\sum_{\br'}}
\lambda_x(\br_1,\br') A(\br', \br_2'; \omega) ,
\label{Qdef}
\end{align}
whose trace and integral over $\omega$ gives $2\lambda_x$. The quantity $Q_x$ may be regarded 
as a decomposition of $\lambda_x$ in terms of quasiparticle energies $\omega$. 
Eq. \eqref{Pi} for the polarization becomes
\begin{align}
\begin{split}
\Pi_{xx}(i\omega) &= 
- \frac{1}{N} \mathop{\sum_{\br_1\subcomma \br_2}} \int dE\: dE' \: 
\frac{f(E) - f(E')}{i\omega - (E'-E)}
\\ &\mytrace \left[ Q_x(\br_1,\br_2; E) Q_x(\br_2,\br_1; E') \right] .
\end{split}\label{Pi_from_Qx} 
\end{align}
The ZSA is implemented in the disordered system
by approximating $Q_x$ by a form which is valid for low-lying quasiparticle
energies $E,E'$.

To find such an approximation, 
it is instructive to view $\lambda_x(\br_1,\br_2)$ as the kernel of an integral operator, which we 
call the ``vertex operator''. To illustrate, we rewrite \eqref{lambda} as 
\begin{align}
\lambda_x(\br_1,\br_2) = \iint \!\! 
\frac{d^2(ka)}{(2\pi)^2} 
\,2 t \sin(k_x a)
\,e^{i \mathbf{k} \cdot \br_1}
\,(e^{i \mathbf{k} \cdot \br_2})^* .
\label{lambda_op}
\end{align}
The vertex operator thus has eigenvalues $2 t \sin(k_x a)$ and plane-wave eigenvectors.
On the other hand, the quasiparticle wavefunctions of the \emph{clean} system
(i.e. $V_{\br} = 0$, $t^{\perp}_{\br} = 0$, and $\Delta_{\br}$ constant) are also plane waves.
The $\delta$-function behavior of conductivity in the clean system arises precisely 
because the vertex function and the Hamiltonian are diagonal in the same basis, i.e. 
the vertex operator commutes with $H$. In the disordered system, $H$ is diagonalized by the 
wavefunctions $\fraku^{(n)}$ and thus does not commute with the vertex operator, making $\sigma_1(\omega)$ regular.

A low-energy effective vertex operator for the clean system is obtained by restricting 
$\mathbf{k}$ in \eqref{lambda_op} to the vicinity of a node $\mathbf{q}$. Under this restriction, 
the eigenvalue is weakly $\mathbf{k}$-dependent.
Thus we make the approximation
\begin{align}
2t \sin(k_x a) \rightarrow \pm v_F/a\sqrt{2},
\label{lambda_approx}
\end{align}
the sign depending on the node.
The leading $\mathbf{k}$-dependent term $(k_x - q_x)/m^*a$ 
leads in the clean system to a correction to \eqref{mainresult} whose relative size we estimate 
as $(2 \eta T / m^* v_F^2)^2$ where $\eta > 1$ is the anisotropy of the linearized nodal dispersion.
For the particular bandstructure of \eqref{Hamiltonian}, 
$m^*v_F^2 = 4 t \sin q_xa \tan q_xa$, which is nominally of order $4t$ but diverges rapidly as
half-filling is approached. Furthermore, $\eta T_c \ll 4t$ in underdoped cuprates, making
\eqref{lambda_approx} an excellent approximation.

The approximate vertex operator arising from \eqref{lambda_approx} is, 
within the subspace of a single nodal region, simply a constant multiple of the identity operator.
This powerful simplification may be applied to the disordered system given two assumptions.
We first assume that the low-energy wavefunctions in the disordered system 
are linear combinations of low-energy wavefunctions in the clean system. 
We thus neglect the contribution of low-energy states which may arise from the 
mixing of anti-nodal states.\cite{SheehyDOS} Since the anti-nodes are connected by wavevectors
of order $a^{-1}$, such low-energy states should constitute a negligible portion
of the spectral density when $\Lambda a \ll 1$. 
Second, we assume that the extended impurities produce negligible internodal scattering. 
Degenerate plane-wave states from different nodes will hybridize and split by an energy 
of order $|V'|$, where $V'$ is a typical matrix element for scattering between nodes. 
Assuming the Fourier spectrum of $V_{\br}$ to be exponential with width $\Lambda$, we estimate
$|V'| \sim V_0 e^{-|\delta q|/\Lambda}$ where $\delta q$ is the relative momentum between 
nodes and $V_0$ is the RMS value of the potential $V_{\br}$.
\endnote{Generally, $\Delta$ and $t^{\perp}$ will have spatial variation as well as $V$. In this case, 
$V_0$ characterizes the largest of the three fluctuations.}
Already $|V'| \ll V_0$ 
when $\Lambda^{-1}$ is comparable to just a few lattice spacings. We assume that relevant experiments are
in the regime $T \gg |V'|$, 
so that these hybridized states are essentially degenerate.\endnote{
Beyond the ZSA, the splitting of these states governs the broadening of 
the $\delta$-function in conductivity.}
We may then choose
a basis for the low-energy sector of the disordered system in which each state resides near  
a \emph{single} node.  

Thus, in the regime $|V'| \ll T \ll 4t/\eta$, it becomes trivial to express the vertex operator in 
the quasiparticle basis of the disordered system. We denote the eigenvalues 
as $\lambda_x^{(n)} = \pm v_F/a\sqrt{2}$, the sign depending on the node of state $n$. 
Then \eqref{Qdef} becomes
\begin{align}
Q_x \rightarrow &\frac{1}{N} \mathop{\sum_n}  
\lambda_x^{(n)} 
\left[ \delta(\omega - \epsilon_n) \fraku_{\beta}^{(n)}(\br_1) \fraku_{\alpha}^{(n)\,*}(\br_2)
\right. \label{Qapprox} \\ &- \left.
\nonumber
\delta(\omega + \epsilon_n) \varepsilon_{\alpha\alpha'}\varepsilon_{\beta\beta'}
\fraku_{\alpha'}^{(n)}(\br_2) \fraku_{\beta'}^{(n)\,*}(\br_1) \right] ,
\end{align}
which may be regarded as the precise statement of the ZSA.
Since the time reversal transformation on wavefunctions is 
$\fraku^{(n)}_{\alpha}(\br) \rightarrow \varepsilon_{\alpha\alpha'}\fraku^{(n)\,*}_{\alpha'}(\br)$, 
the second term in \eqref{Qapprox} corresponds to the time-reversal of the first and acquires a 
sign change.
The orthogonality relations \eqref{frakunitarity} allow us to evaluate
\begin{align}
\frac{1}{N} &\mathop{\sum_{\br_1\subcomma \br_2}}
\mytrace \left[ Q_x(\br_1, \br_2; E) Q_x(\br_2,\br_1; E') \right]
 = 
\\ \nonumber &
\frac{1}{2}\left(\frac{v_F}{a}\right)^2 \delta(E-E') 
\frac{1}{N} \mathop{\sum_n}[\delta(E - \epsilon_n) + \delta(E + \epsilon_n)] .
\end{align}
Substituting into \eqref{Pi_from_Qx} leads directly to 
main result \eqref{mainresult}, with the density of states per site per spin $\dos$ given by 
\begin{align}
\dos(\omega) &\equiv \frac{1}{2N}\mathop{\sum_{n}} 
[\delta(\omega - \epsilon_n) + \delta(\omega + \epsilon_n)] .
\end{align}

We note that \eqref{mainresult} can also be obtained\cite{apy} using the 
diagram method\cite{AGD} for disorder-averaged correlation functions.  In this context, zero-angle scattering
at the nodes allows vertex correction diagrams to be summed exactly and expressed in terms of the single-particle 
Green function. This simplification is formally analogous to the diagrammatic Ward-Takahashi 
identity\cite{PeskinQFT} of quantum electrodynamics, by which the longitudinal component of the 
interacting vertex function may be expressed in terms of the interacting Green function. In our context, 
however, \eqref{mainresult} has nothing to do with electromagnetic gauge invariance, but arises 
instead from the fact that i) the Hamiltonian \eqref{Hamiltonian} is noninteracting and ii) the 
effective vertex function for the low-energy states has trivial structure, as we have argued above. 

In ordinary metals forward scattering leads to a reduction of the 
transport scattering rate 
$1/\tau_{\rm tr}$ compared to the single-article scattering 
rate $1/\tau$. The result presented above can be viewed as an extreme 
example of this effect in a system of nodal fermions where $1/\tau$ is finite
but $1/\tau_{\rm tr}$ is reduced essentially to zero. This illustrates the 
singular nature of the vertex corrections that have been summed to obtain   
\eqref{mainresult}. It is also interesting to note that the superfluid density, 
ordinarily considered a transport quantity, depends only
on the single-particle scattering rate in this case.

We have derived the remarkably simple result \eqref{mainresult}, 
which relates the normal fluid density to the quasiparticle 
density of states in a system disordered with extended impurities.
The normal fluid density may be computed in the ZSA 
in spite of the fact that,
as in a clean superconductor, 
the real part of conductivity becomes a $\delta$-function
at zero frequency. This derivation proceeds directly from the 
quasiparticle wavefunctions $\fraku^{(n)}$ and energies $\epsilon_n$ of Hamiltonian \eqref{Hamiltonian},
showing that \eqref{mainresult} holds for individual realizations of the disorder and thus circumvents the 
considerable complications of vertex corrections. 
It therefore allows for a model-independent inversion of $n_n(T)$ from 
experiment to obtain the density of states. To theoretically predict $\dos(E)$ and $\sigma_1(\omega)$,
however, requires a more specific, model-dependent calculation.  Eq. \eqref{mainresult} may be applicable to 
high-purity single crystals of underdoped YBCO,
in which slowly varying potentials from interlayer dopant disorder 
dominate the low energy scattering. 

We acknowledge A. J. Berlinsky, D. A. Bonn, T. P. Davis, C. Kallin, and 
M. Schechter for helpful discussions. This work was supported by NSERC,
CIAR, and the A. P. Sloan Foundation. The authors are indebted to Aspen Center 
for Physics where this work was initiated.

\bibliographystyle{apsrev_vc}
\bibliography{vc}

\end{document}